\documentclass[prb,twocolumn,superscriptaddress,showpacs,floats,floatfix]{revtex4}

\usepackage{epsfig,dcolumn,amsmath,latexsym}

\begin{document}

\newcommand{\Go}{\ensuremath{G_0}}
\newcommand{\IV}{\mbox{$I$-$V$}}
\newcommand{\VR}{\ensuremath{dV/dt}}
\newcommand{\etal}{\textit {et al.}\/}
\newcommand{\figref}[1]{Fig.~\ref{fig.#1}}
\renewcommand{\eqref}[1]{Eq.~(\ref{eq.#1})}
\newcommand{\tableref}[1]{Table~\ref{table.#1}}

\title{Conductance of single-atom platinum contacts:\\ Voltage-dependence of the conductance histogram}

\preprint{7}

\author{S.~K.~Nielsen}
\affiliation{Interdisciplinary Nanoscience Center (iNANO), CAMP
and  Department of Physics and Astronomy, University of Aarhus,
DK-8000 Aarhus, Denmark}

\author{Y.~Noat}
\thanks{Present address: Groupe de Physique des Solides, Campus Jussieu tour 23, 2 Place Jussieu,
75251 Paris cedex 05 France} \affiliation{Kamerlingh Onnes
Laboratory,Universiteit Leiden, Box 9504, 2300 RA Leiden, The
Netherlands}

\author{M.~Brandbyge}
\affiliation{
Mikroelektronik Centret (MIC), Technical University of Denmark, Bldg. 345E, DK-2800 Lyngby, Denmark}

\author{R.~H.~M.~Smit}
\affiliation{Kamerlingh Onnes Laboratory,Universiteit Leiden, Box
9504, 2300 RA Leiden, The Netherlands}

\author{K.~Hansen}
\affiliation{Interdisciplinary Nanoscience Center (iNANO), CAMP
and  Department of Physics and Astronomy, University of Aarhus,
DK-8000 Aarhus, Denmark}

\author{L.~Y.~Chen}
\affiliation{Kamerlingh Onnes Laboratory,Universiteit Leiden, Box
9504, 2300 RA Leiden, The Netherlands}

\author{A.~I.~Yanson}
\thanks{Present address: Dept.\ of Physics, 510 Clark Hall, Cornell University, Ithaca, NY 14853 }
\affiliation{Kamerlingh Onnes Laboratory,Universiteit Leiden, Box
9504, 2300 RA Leiden, The Netherlands}

\author{F.~Besenbacher}
\affiliation{Interdisciplinary Nanoscience Center (iNANO),  CAMP
and Department of Physics and Astronomy, University of Aarhus,
DK-8000 Aarhus, Denmark}

\author{J.~M.~van~Ruitenbeek}
\email[Corresponding author ]{ruitenbe@Phys.LeidenUniv.nl}
\affiliation{Kamerlingh Onnes Laboratory,Universiteit Leiden, Box
9504, 2300 RA Leiden, The Netherlands}

\date{\today}

\begin{abstract}
The conductance of a single-atom contact is sensitive to the
coupling of this contact atom to the atoms in the leads. Notably
for the transition metals this gives rise to a considerable spread
in the observed conductance values. The mean conductance value and
spread can be obtained from the first peak in conductance
histograms recorded from a large set of contact-breaking cycles.
In contrast to the monovalent metals, this mean value for Pt
depends strongly on the applied voltage bias and other
experimental conditions and values ranging from about 1\,\Go\ to
2.5\,\Go\ ($\Go=2e^2/h$) have been reported. We find that at low
bias the first peak in the conductance histogram is centered
around 1.5\,\Go. However, as the bias increases past 300~mV the
peak shifts to 1.8\,\Go. Here we show that this bias dependence is
due to a geometric effect where monatomic chains are replaced by
single-atom contacts, where the former are destabilized by the
electron current at high bias.
\end{abstract}


\maketitle


The conductance of atomic-sized contacts (ASCs) is ballistic and
can be written in terms of eigenchannels of the contact. In the
limit of a single-atom contact the number of eigenchannels is
governed by the number of valence orbitals of the atom.
\cite{scheer98} The simplest description, with a single
conductance channel, applies for monovalent metals (Au, Na, ...),
while for $sp$ metals three channels contribute, and for
transition metals with open $d$ shells five channels contribute to
the conductance. In general the transmission for each of the
channels is smaller than 1, but one often finds that the single
channel for the monovalent metals is nearly perfectly transmitted
giving a conductance of almost one conductance quantum,
$\Go=2e^2/h$, where $e$ is the electron charge and $h$ is Planck's
constant. The five channels for the transition metals add up to a
total conductance that sensitively depends on the coupling of the
atom to its neighboring atoms in the leads and usually ranges
between 1.5\,\Go\ and 3.5\,\Go. This can be judged from the
position and width of the first peak in so-called conductance
histograms. For a recent review we refer to
Ref.\,\onlinecite{Agrait02}.

The concept of conductance histograms was introduced to
investigate possible conductance quantization in metallic
contacts.\cite{Olesen95a,Krans95b} The histograms are constructed
from digitized traces of thousands of cycles of breaking (or
making) ASCs. These traces are projected onto the conductance
axis, and in this way preferred conductance values become visible
as peaks. Only for monovalent metals, including the noble
metals\cite{Brandby95a,Ludoph00a} and the alkali
metals,\cite{Krans95b,Ludoph00a} some form of quantization is
observed as evidenced by prominent peaks in the histograms close
to multiples of the conductance quantum. For most other metals
there is no indication of quantization in the conductance
histograms.\cite{Ludoph00a,Agrait02} There is usually a single
peak at low conductance in the histograms that indicates the
preferred conductance of the single-atom contacts.\cite{Ludoph00a}
For most metals this first peak, when measuring in a clean
environment, is fairly reproducible and insensitive to the level
of the applied bias voltage. For Au, which has been most studied,
it has been shown that the conductance of the first peak is bias
independent up to about 2\,V, which can be seen in the statistical
average represented by conductance histograms measured for various
bias voltages\cite{SakaiAu} and from individual curent-voltage
(\IV) curves.\cite{Hansen00a,Nielsen02a}

\begin{figure}[!t]
\begin{center}
\epsfig{width=6cm,figure=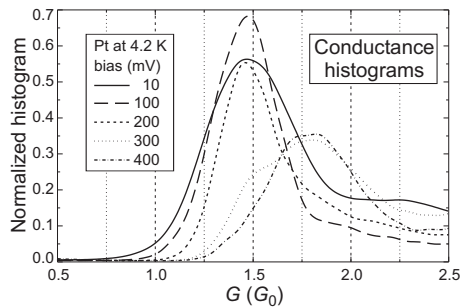} \caption{\small Pt conductance
histograms, each compiled from 3000 conductance traces of
\textit{breaking} ASCs recorded at 4.2\,K under cryogenic vacuum
in the MCBJ. The histograms are measured in succession on the same
sample as the bias increases from 10 to 400\,mV and displays a
clear dependence on the bias. (The samples are always made from a
Pt wire with a purity of 99.999\% and a diameter of 0.1\,mm). The
histograms have been normalized.\cite{Norm} }
\label{fig.bias-dependence}
\end{center}
\end{figure}

Pt is a marked exception, and widely different results have been
presented for this metal. The reported low bias position of the
first peak varies between 1.0\,\Go\ and
2.5\,\Go.\cite{Krans93a,Olesen95a,Sirvent96b,Yuki00a,Smit01a,Smit02b}
Bias dependences of Pt conductance histograms have also been
reported,\cite{Yuki00a} where a low bias peak at 1.0\,\Go\ was
replaced by a peak at 1.7\,\Go\ at higher biases. Recently it has
been shown that at low bias the first peak is centered around
1.5\,\Go\ (as in \figref{bias-dependence}), and that a peak at
1.0\,\Go\ can be caused by the presence of hydrogen molecules that
act as the final bridge before the contact breaks.\cite{Smit02b}

In this paper we investigate the bias dependence of Pt conductance
histograms, compiled from thousands of ASCs formed using the
mechanically controllable break-junction
(MCBJ)\cite{Krans93a,Agrait02} at liquid helium
temperature~(4.2\,K) in a cryogenic vacuum environment. The
conductance measured is the linear conductance $G=I/V$. A small
but significant shift in the position of the first histogram peak
from 1.5\,\Go\ to 1.8\,\Go\ is observed when the bias increases
past 300\,mV\@. Contrary to what would have been expected from
measured \IV\ curves of Pt,\cite{Nielsen02a} where the conductance
decreases with voltage, the shift is towards higher conductance.
This indicates that the shift is not caused by an electronic
effect. We present evidence that the shift marks a geometric
transition point where single-atom Pt contacts replace the
monatomic chains known to exist for Pt.\cite{Smit01a}


The bias dependence is illustrated by the conductance histograms
shown in \figref{bias-dependence}. For the curve recorded at a low
bias of 10\,mV the first peak is found close to 1.5\,\Go.
The conductance histogram recorded at 100\,mV is very similar to
the one at 10\,mV with a broad peak centered around 1.5\,\Go.
Increasing the bias further to 200\,mV only causes a slight
decrease in the intensity, whereas the position remains unchanged.
However, when the bias increases to 300\,mV, the main part of the
histogram peak switches to be centered around 1.8\,\Go\, although
a broad shoulder can still be seen around 1.5\,\Go. At 400\,mV,
the shoulder has disappeared and the peak center has completely
moved from 1.5\,\Go\ to 1.8\,\Go. Also the peak intensity
decreases considerably for the two high bias conductance
histograms compared with those measured at lower bias. We have
verified that the process is reversible, by lowering the bias back
to zero observing the original histogram with a peak close to
1.5\,\Go\ reemerge.


The peak shift to higher $G$ is very puzzling when compared to the
Pt \IV\ curves presented previously by Nielsen \etal\
\cite{Nielsen02a} It was shown that \IV\ curves of Pt are markedly
non-linear but with a decreasing conductance as the voltage
increases. The decrease in $G$ was found to be proportional to the
voltage squared, resulting only in a limited decrease of the
linear conductance at 300~\,mV which amounts to about $-0.1$\,\Go\
at 500\,mV. The important observation is, however, that the
conductance always decreases and never increases with voltage, the
opposite effect of the one observed with the conductance
histograms presented in \figref{bias-dependence}. This clearly
indicates that the peak shift is not caused by an electronic
effect.


Investigating other possible causes we turned to the formation of
monatomic chains, which was first discovered for
Au.\cite{Ohnishi98a,Yanson98a} Recently it was shown by Smit
\etal\ that also Pt and Ir have this property.\cite{Smit01a} To
demonstrate the formation of monatomic chains, Smit \etal\
recorded plateau-length histograms. It is important to emphasize
the difference between plateau-length
histograms\cite{Yanson98a,Smit01a} and the conductance histograms
used above. The peaks in the conductance histograms reveal the
most probable conductance values obtained for breaking ASCs. A
plateau-length histogram is constructed by measuring the lengths
of the last conductance plateau, that corresponds to a conductor
of a single atom in cross-section, from thousands of conductance
traces of breaking ASCs. By plotting the number of times a given
length occurs, a histogram is obtained with peaks revealing the
typical disruption lengths, at which the monatomic chains tend to
break. The first peak reflects the length of the single-atom
contact and peaks for longer wires indicate the formation of
monatomic chains that break at 2, 3, 4,... atoms in length.

\begin{figure}[!b]
\begin{center}
\epsfig{width=6cm,figure=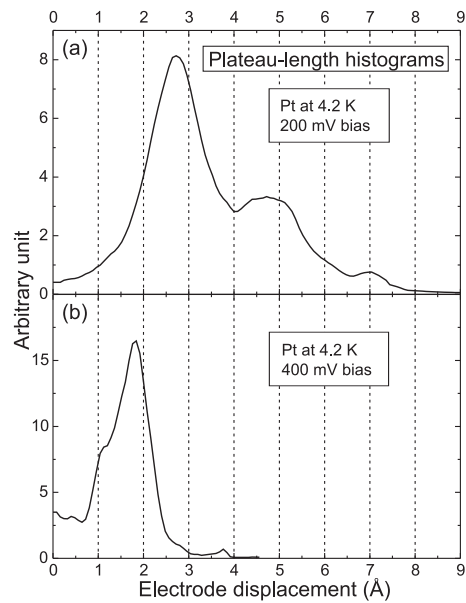} \caption{\small Plateau-length
histograms\cite{Yanson98a,Smit01a} for Pt, each compiled from 2000
breaking ASCs. A bias of (a)~200\,mV and (b)~400\,mV is applied.
The vertical axis gives a measure for the frequency with which a
given chain length occurs. } \label{fig.length-histogram}
\end{center}
\end{figure}

In \figref{length-histogram} we show two plateau-length histograms
recorded at biases of (a) 200\,mV and (b) 400\,mV, respectively.
At 200\,mV bias, the plateau-length histogram is very similar to
the one presented by Smit \etal\ at 10\,mV bias.\cite{Smit01a}
Three clear peaks are visible, the first corresponding to a
single-atom contact (a `one atom chain'), and the next peaks
represent chains with two and three atoms, respectively. Thus, we
conclude that monatomic chains still form at a bias as high as
200\,mV\@. The three peaks in \figref{length-histogram}(a) are
centered at 2.7, 4.8 and 7.0\,\AA, respectively. The distances
between them are thus \mbox{2.1--2.2\,\AA}, similar to the
2.3\,\AA\ found by Smit \etal\ at 10\,mV bias,\cite{Smit01a}
within the accuracy of $\sim$10\% in our length calibration. The
position of the first peak in \figref{length-histogram}(a) differs
from this peak-to-peak distance as can be expected. It reflects
the elastic stretching of the banks and the bonds to a single
bridging atom at the verge of breaking.

The presence of only one single peak in the 400\,mV plateau-length
histogram of \figref{length-histogram}(b), centered at 1.9\,\AA,
proves that chains no longer form at this high bias. The peak does
not coincide with the corresponding first peak in the 200\,mV
plateau-length histogram. This difference is most likely due to
the higher bias causing the single-atom contact to break at a
smaller strain than at low bias (see below).

\begin{figure}[!t]
\begin{center}
\epsfig{width=6cm,figure=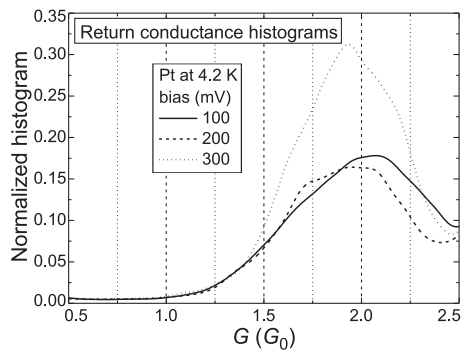} \caption{\small Return
conductance histograms of Pt. Each histogram is compiled from 2000
conductance traces and is measured in succession on the same
sample while the bias increases from 100 to 300\,mV in 100\,mV
steps. The measurements are performed by \textit{forming} the ASCs
with the MCBJ. The histograms have been normalized.\cite{Norm} }
\label{fig.return-histogram}
\end{center}
\end{figure}

From the plateau-length histograms we find that the formation of
atomic chains is inhibited above the bias voltage for which also
the shift in the first peak in the conductance histogram is
observed (300\,mV). This suggests that the chain formation affects the peak
position in the conductance histograms. To test this we
recorded conductance histograms for curves measured while
returning to contact from the vacuum tunneling regime, which we
will refer to as return histograms. The conductance histograms
presented previously have all been obtained from the conductance
traces of \textit{breaking} ASCs. When breaking the contacts,
monatomic chain formation may occur. Instead, when we obtain the
ASCs by returning the electrodes back into contact, chains cannot
form. Measuring the conductance traces while \textit{forming}
ASCs, results in the return histograms presented in
\figref{return-histogram}. From these return histograms it is clear that the first peak is
centered close to 2\,\Go, independent of bias. It thus seems that
the peak shift occurs when the chain formation is inhibited, since
the peak is located at an even higher conductance, independent of
bias, when no chains can form.

\begin{figure}
\begin{center}
\epsfig{width=7cm,figure=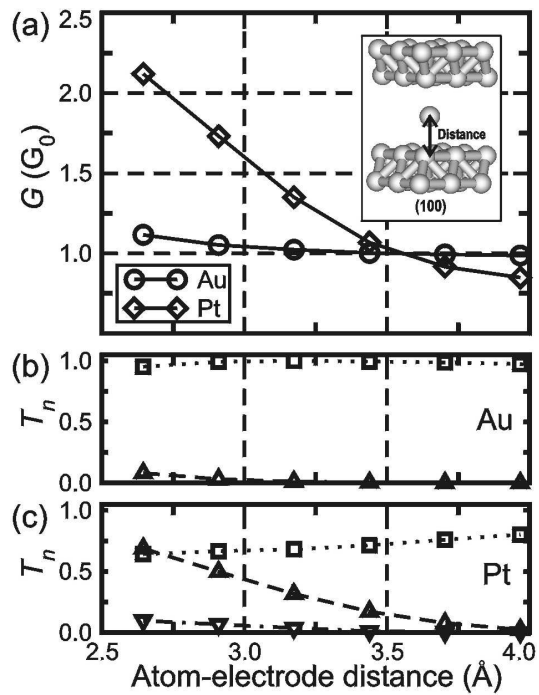} \caption{\small (a)
Calculated conductance $G$ vs atom-electrode distance at zero
voltage for a symmetric single-atom Au or Pt contact configuration
consisting of a single atom between two Au or Pt(100) surfaces
(inset). Note that the contacts in practice will become unstable
beyond a distance of about 3\AA. (b) and (c) The conductance
decomposed into eigenchannel transmissions for Au and Pt. The
dotted line corresponds to the non-degenerate channel consisting
of $s$ and $d_{z^2}$ orbitals, the dashed line corresponds to the
two degenerate channels with the $d_{zx}$ and $d_{yz}$ orbitals,
while the dot-dashed corresponds to the single channel with
$d_{x^2-y^2}$. } \label{fig.calculation}
\end{center}
\end{figure}

For Au the conductance of a single-atom contact is
indistinguishable from that of monatomic chains and we had
expected a similar result for Pt, but the present results suggest
otherwise. To get further insight into these findings, we have
used the \mbox{\sc Transiesta} program\cite{TransSiesta} to
calculate the conductance and eigenchannel transmissions of
single-atom Au and Pt contacts [inset of \figref{calculation}(a)]
as the atom-electrode distance increases. The results are shown in
\figref{calculation}. The method is based on density functional
theory and takes the voltage bias and current explicitly into
account in the self-consistent calculation of electronic density
and potential. Our calculations show that, contrary to Au for
which the conductance remains stable around 1\,\Go, single-atom Pt
contacts display a strong variation in the zero-bias conductance
with a decrease from 2.1\,\Go\ to 1.1\,\Go\ as the distance
increases from 2.65 to 3.5\,\AA. This variation is in accordance
with the reported broad histogram peak centered around
(1.5--2)\,\Go\ for Pt.\cite{Sirvent96b,Smit01a,Smit02b} These
results can help explain the behavior leading to the three types
of histograms presented above.

The histogram with a peak at 2\,\Go\ (\figref{return-histogram})
is produced by single-atom contacts. In this case the electrodes
are moved towards each other such that the atom-electrode distance
is reduced to a minimum. The peak at 1.8\,\Go\ in the high-bias
histograms (\figref{bias-dependence}) also results from
single-atom contacts, but in this case the breaking of the wire
leads to a larger atom-electrode distance, and thus a lower
conductance (\figref{calculation}). The peak at 1.5\,\Go\  in the
low-bias histograms of \figref{bias-dependence} is due to the
frequent occurrence of atomic chains, which are only stable for
sufficiently low bias voltages, below 300\,mV. The lower average
conductance for atomic chains is suggested by the physics
described in \figref{calculation}. As the single atom is
positioned farther away from the banks the overlap of its orbitals
with those in the electrodes is reduced leading to the observed
decrease in the conductance. For monatomic chains the number of
states overlapping with the central atoms in the chains is also
reduced with respect to the single-atom case, leading to a similar
reduction of the conductance. The suppression of chain formation
at higher bias is partly due current-induced embrittlement, but
mainly results from heating of the atomic degrees of freedom by
the current \cite{Todorov01,Smit03}.


We can now attempt to classify the various results found in the
literature. A first peak in the conductance histogram close to
1\,\Go\ was found in room temperature experiments under ultra-high
vacuum,\cite{Olesen95a} or under an atmosphere of N$_2$+5\%H$_2$
gas.\cite{Yuki00a} For the latter experiment the peak was seen to
move to 1.7\,\Go\ at elevated bias. For all experiments performed
under cryogenic vacuum and at a voltage bias below 100\,mV (Refs.
\onlinecite{Krans93a,Sirvent96b,Smit01a,Smit02b} and this work)
the first peak is found at (1.5--1.6)\,\Go. However, when
hydrogen is intentionally introduced into the vacuum system a peak
near 1\,\Go\ appears, which has been interpreted as being due to
the formation of a conducting hydrogen bridge forming the last
contact.\cite{Smit02b} Assuming this mechanism is still effective
at room temperature the results by Yuki \etal\cite{Yuki00a} find a
natural explanation. Trace amounts of hydrogen in the UHV system
at room temperature may similarly explain the observation by
Olesen \etal\cite{Olesen95a}

The remaining variation in the position of the first peak in the
Pt conductance histograms can be attributed to a bias-dependence
of the formation of monatomic chains. The shift of the first peak
in the Pt conductance histogram from 1.5\,\Go\ to 1.8\,\Go\ when
the bias increases past 300\,mV, marks a geometric transition
point where monatomic chains are replaced by single-atom
contacts.\cite{NielsenPHD}


We acknowledge financial support from The Center for Atomic-scale
Materials Physics (CAMP) sponsored by the Danish National Research
Foundation and from the EU network ``Bottom up Nanomachines'' (BUN). MB
acknowledges support from the Danish Natural Science Research
Council. YN has been supported by a Marie Curie fellowship of the
European Community under contract number HPMF-CT-1999-00196.


\begin{thebibliography}{10}
\bibitem{Norm}
The conductance histograms in \figref{bias-dependence} and
\ref{fig.return-histogram} have been normalized by the area under
the curve in the conductance range from (0--10)\,\Go. Since the
figures show a zoom to (0.5--2.5)\,\Go\ the area in the figures
are not the same.
\end{thebibliography}
\end{document}